\documentclass[12pt,preprint]{aastex}
\usepackage{natbib}
\usepackage{amsmath}
\usepackage{times}
\usepackage{graphicx}
\usepackage{epstopdf}
\usepackage{textcomp}
\usepackage{color}

\def\epsl{Earth Planet. Sci. Lett.}%

\newcommand{\myemail}{ewirstrom@gmail.com}

\shorttitle{Spin-state dependent $^{15}$N fractionation}
\shortauthors{Wirstr\"om et al.}

\begin{document}


\title{Isotopic Anomalies in Primitive Solar System Matter: Spin-state Dependent Fractionation of  Nitrogen and Deuterium in Interstellar Clouds}


\author{Eva S. Wirstr\"om\altaffilmark{1,2}, Steven B. Charnley\altaffilmark{1}, Martin A. Cordiner\altaffilmark{1,2} and Stefanie N. Milam\altaffilmark{1}}
\affil{$^1$Astrochemistry Laboratory and The Goddard Center for Astrobiology, \\NASA Goddard Space Flight Center, Greenbelt, MD 20770, USA}
\email{\myemail}


\altaffiltext{2}{Also at Department of Physics, The Catholic University of America, Washington, DC 20064, USA}


\begin{abstract}
{
Organic material found in meteorites and interplanetary dust particles is enriched in D and $^{15}$N. This is consistent with the idea that the functional groups carrying these isotopic anomalies, nitriles and amines, were formed by ion-molecule chemistry in the protosolar nebula.
Theoretical models of interstellar fractionation at low temperatures predict large enrichments in both D and $^{15}$N and can account for the largest  isotopic enrichments measured in carbonaceous meteorites. However,   more recent measurements have shown that,  in some primitive samples, a large $^{15}$N enrichment does not correlate with one in  D, and that some D-enriched primitive material displays little, if any,  $^{15}$N enrichment.
By considering the spin-state  dependence in ion-molecule reactions involving the ortho and para forms of  H$_2$,    we show that ammonia and related molecules can  exhibit such a wide range of fractionation for both $^{15}$N and D in dense cloud cores. We also show that while the nitriles, HCN and HNC, contain the greatest   $^{15}$N enrichment, this is  not expected to correlate with extreme D enrichment.  
 These calculations therefore support the view that Solar System $^{15}$N and D isotopic anomalies have an interstellar heritage. We also  compare our results to existing astronomical observations and briefly discuss future tests of this model.
}

\end{abstract}




\section{Introduction}
 
The isotopic enrichments measured in the primitive organic matter found in comets, interplanetary dust particles (IDPs) and meteorites  \citep{Remusat06,Floss06,Aleon10}, probably had their chemical origin in a low-temperature environment that faciliated ion-molecule isotopic exchange reactions, such as the dense interstellar medium (ISM) or the outer protosolar nebula \citep{MummaCharnley11,Messenger03}.

The enhanced D/H ratios seen in some primitive material \citep{Messenger00,Robert03,Messenger03,Keller04,Hartogh11} almost certainly originated in low-temperature environments.  Deuterium  enhancement is initiated by the exothermic ion-molecule exchange reaction \citep{Millar89} going to the right 
\begin{equation}\label{HDReac}
{ \rm{H_3^+~~+~~ HD~\rightleftharpoons~ H_2D^+ ~~+~~H_2~~+~~170~K} }  
\end{equation}

 In dense cores where CO and other heavy molecules have condensed as ices on to dust grains (i.e. become depleted), H$_2$D$^+$  molecules can continue to react with HD,  producing HD$_2^+$ and D$_3^+$  \citep{Roberts03}.  This produces very high levels of  deuteration in the remaining gaseous species as well as, through the resulting high atomic D/H ratios, in molecular ice mantles \citep[e.g.][]{Ceccarelli07}.  A similar  D fractionation chemistry can also occur in the cold  regions of protoplanetary disks \citep{Willacy07}.

In both Jupiter Family and Oort Cloud comets,  $\rm ^{14}N/^{15}N $ ratios are measured ($\sim 130-170$) for both CN and HCN \citep{Hutsemekers08,BockeleeMorvan08} that are significantly enhanced relative to both the terrestrial and protosolar values \citep[272 and 440 respectively,][]{Marty12}. 
 Primitive materials can  exhibit large bulk $^{15}$N enhancements ($\rm ^{14}N/^{15}N \sim 180-300$), and  high  spatial resolution measurements further show that they also contain distinct `hotspots' which exhibit the greatest D and $^{15}$N enhancements known ($\rm ^{14}N/^{15}N <100$) \citep{Messenger00,Aleon03,Floss06,Busemann06,Briani09b,Aleon10}.

Theoretical models of nitrogen isotopic chemistry in molecular cloud cores, where  gaseous molecules are being depleted by accretion on to dust grains, can reproduce the range of $\rm ^{14}N/^{15}N$ ratios found in primitive matter \citep[$\sim 50-280$, ][]{Aleon10} through exchange reactions between $^{15}$N atoms and molecular ions such as $^{14}$N$_2$H$^+$ and HC$^{14}$NH$^+$  \citep{CharnleyRodgers02,RodgersCharnley08,RodgersCharnley08_ApJ}.   These models predict that N$_2$, NH$_3$, HCN, CN and HNC  molecules should possess the  greatest $^{15}$N enrichments.  There are two $^{15}$N-fractionating pathways \citep{RodgersCharnley08_ApJ}: a slow one to N$_2$ and ammonia ($\sim 10^6$ years) and more rapid one to
HCN and other nitriles ($\sim 10^5$ years).

Recent, albeit sparse,  astronomical observations of starless cores indicate that interstellar  ${\rm  NH_3}$ and ${\rm  N_2}$ are not as enriched as predicted for the ISM or as measured in the Solar System:  ${\rm ^{14}NH_3/^{15}NH_3} \approx 334$ \citep{Lis10} and $\rm ^{14}N_2H^+/^{14}N^{15}NH^+   \approx 446$  \citep{Bizzocchi10}.

However, the  most pressing  problem for a direct ISM-Solar System isotopic connection concerns the fact  that, while $^{15}$N  and D meteoritic hotspots do seem to correlate in some samples \citep[e.g.][]{Aleon03}, they clearly do not in others \citep[e.g.][]{Busemann06,Gourier08}.  Yet, as has been noted by several authors \citep{Alexander08,Briani09,Bonal10,Aleon10}, the low-temperature interstellar environments most conducive to producing  large  $^{15}$N enrichments, should also produce concomitantly enormous molecular  D/H ratios (e.g. in ammonia or HCN). Thus, one would expect $^{15}$N  and D hotspots to {\it always} be spatially correlated, contrary to what is seen. This presents a serious challenge for ion-molecule fractionation chemistry.

 
In cold interstellar environments,  nuclear spin states can be an important factor in determining chemical reaction rates \citep[e.g.][]{Flower06}. The major interstellar collision partner,   molecular hydrogen,  has a 170.5~K difference in zero point energy between the $para$ (antiparallel spins) and the $ortho$ (aligned nuclear spins) forms;   this allows some endoergic reactions to proceed, even at low temperatures, if a substantial fraction of the H$_2$ is in the higher energy form, $o$-H$_2$.  
\citet{Pagani11} demonstrated that a high H$_2$ ortho-to-para ratio (OPR) therefore inhibits (`poisons') the  production of H$_2$D$^+$ in reaction (1), suppressing the overall deuteration. 

The fractionation of $^{15}$N may also be influenced by the OPR of H$_2$. 
 Ammonia formation is initiated by the production of N$^+$ from N$_2$ by He$^+$, which then reacts  in
\begin{equation} \label{amminitReac}
{\rm N^+~+~H_2} ~\longrightarrow~ {\rm NH^+~+~H},
\end{equation}
 followed by a sequence of ion-molecule reactions with H$_2$ through to NH$_4^+$ and a final  electron dissociative recombination.  
Reaction \ref{amminitReac} has an activation energy of $\lesssim$200 K \citep{Gerlich93} which, at low temperatures,  can be overcome by the internal energy of \textit{o}-H$_2$. Thus, ammonia formation and the related $^{15}$N fractionation in dark clouds are dependent on the H$_2$ OPR \citep{LeBourlot91}, a distinction not made in earlier models \citep[e.g.][]{RodgersCharnley08_ApJ}.  A recent re-assessment of the OPR dependence in the original experimental data by \citet{Dislaire12} indicates that  previous work has  overestimated the low-temperature rate coefficient for reaction \ref{amminitReac}, involving  \textit{o}-H$_2$,  by almost three orders of magnitude (cf. Le Bourlot 1991).   Thus, the precise abundance of \textit{o}-H$_2$ could  play a pivotal role in producing a diverse range of D-$^{15}$N fractionation in interstellar precursor molecules.

In this Letter we quantify the effect the H$_2$ OPR has on interstellar $^{15}$N fractionation, and demonstrate that this can account for the isotopic anomalies observed in primitive Solar System materials.

\section{Model}
We consider the chemical evolution of the central regions of a static  prestellar core \citep{CharnleyRodgers02} with a gas density $n$(H$_2$)=10$^6$ cm$^{-3}$, a temperature of 10~K and a visual extinction of $A_{\rm V}$$>$10 magnitudes. Cosmic ray ionization occurs, at a rate of $\zeta $=3$\times$10$^{-17}$ s$^{-1}$.
The chemical model used here is based on \citet{RodgersCharnley08,RodgersCharnley08_ApJ}\footnote{We do not consider the neutral-neutral isotope exchange reactions proposed by \citet{RodgersCharnley08_ApJ}} and Fig.~\ref{15NnetworkFig} shows the most important reactions for $^{15}$N fractionation.
Helium, carbon, oxygen and nitrogen are included at elemental abundances  given by \citet{SavageSembach96}.  We are interested in cores that are just about to form protostars and so we assume that  all  carbon is initially bound up in CO, with the remaining oxygen in atomic form, and the nitrogen is partly atomic with 50\% in molecular and atomic form respectively. The elemental $^{14}$N/$^{15}$N ratio is assumed to be 440, with the same fraction of $^{15}$N in atomic form initially, $^{15}$N/$^{15}$N$^{14}$N=1.

   

The gas-grain interaction is modelled according to \citet{Charnley97}: all
neutral gas-phase species, except H$_2$, He, N, and N$_2$, stick and freeze out onto grains upon collision. 
No grain surface chemistry or desorption is considered, except for  sticking of  H atoms (with an efficiency of 0.6)  followed by reaction to produce  H$_2$ molecules which are ejected upon formation (see Sect~\ref{opSect}).

\subsection{Inclusion of spin-state reactions} \label{opSect}
The reaction network has been expanded to contain ortho and para forms of H$_2$, H$_2^+$, and H$_3^+$, and couples 279 gas-phase species by 4420 reactions. Both spin types have been considered equally likely to react with a given species, with the exception of reaction (\ref{amminitReac}), where new rates of \citet{Dislaire12} for ortho and para H$_2$, 
$k_{\ref{amminitReac},o}$=4.20$\times$10$^{-10} (T/300)^{-0.17} {\rm exp}(-44.5/T)$ ~cm$^3$\,s$^{-1}$ and $k_{\ref{amminitReac},p}$=8.35$\times$10$^{-10} {\rm exp}(-168.5/T)$~cm$^3$\,s$^{-1}$, respectively, have been adopted. Other temperature-dependent H$_2$ reactions have energy barriers that are too high for the ground-state energy difference between the spin types to have a significant effect at 10 K.

Initially the H$_2$ OPR is set to the high temperature statistical equilibrium value of 3. 
Measurements in the diffuse ISM by \citet{Crabtree11} are consistent with an H$_2$ OPR around 1; we find adopting this lower value has little effect on our results.
The reaction 
\begin{equation} \label{opH2Reac}
{\rm H^+~+~{\it o-}H_2} ~\longrightarrow~ {\rm H^+~+~{\it p-}H_2}, 
\end{equation}
is exothermic by 170.5 K and so $o$-H$_2$ is efficiently converted into $p$-H$_2$ at low temperatures. We assume that the H$_2$ formed and ejected from grains has an OPR of 3;   this ensures a steady-state OPR is reached. Reaction rates for conversion between ortho and para species, as well as for their interaction with electrons and cosmic rays, have been implemented from \citet{Walmsley04}. Branching ratios for $o/p$-H$_3^+$ formation from H$_2^+$ and H$_2$, and rates for H$_3^+$ interaction with CO and N$_2$, are taken from \citet{Pagani09}. Furthermore, all reactions involving proton transfer from H$_3^+$ to neutral molecules have been assigned branching ratios according to \cite{Oka04} and no spin conversion is assumed to take place in charge exchange reactions involving H$_2^+$.
 Apart from these special cases, reactions are assumed to produce ortho and para versions of H$_2$, H$_2^+$, and H$_3^+$ with equal probability.

\subsection{Other reactions} \label{He+Sect}
Since nitrogen fractionation is initiated by He$^+$ reactions  (see Fig.~\ref{15NnetworkFig}), the He$^+$ chemistry dictates the time-scale of fractionation. In particular, the rate for the reaction 
\begin{equation} \label{HepReac}
{\rm He^+~+~H_2} ~\longrightarrow~ {\rm H^+~+~H~+~He} 
\end{equation}
has previously been assigned widely different values at 10 K, 
ranging from 1.1$\times$10$^{-15}$~cm$^3$\,s$^{-1}$ \citep{RodgersCharnley08_ApJ} to 2.5$\times$10$^{-13}$~cm$^3$\,s$^{-1}$ \citep{Walmsley04}.  
We have adopted the measured low temperature rate $k_{\ref{HepReac}}$=3.0$\times$10$^{-14}$~cm$^3$\,s$^{-1}$ \citep{Schauer89}.

In order to qualitatively compare the temporal patterns in $^{15}$N fractionation to the expected levels of deuteration during core evolution, we also include the chemistry for $ortho$ and $para$ H$_2$D$^+$ in the model since ${\it o-}\rm H_2$ can overcome the barrier for the reverse of reaction (1), i.e.  
\begin{equation}\label{H2DReac}
{\rm H_2D^+~+~{\it o-}H_2}~\longrightarrow~\rm{H_3^+~+~HD}  
\end{equation}
The most important reactions for spin-state dependent $o$/$p$-H$_2$D$^+$ formation and destruction are adapted from \citet{Walmsley04}, and we assume a deuterium abundance corresponding to HD$=1.5\times 10^{-5}$ relative to elemental hydrogen.

\section{Results}
\begin{figure*}
\includegraphics[width=\linewidth]{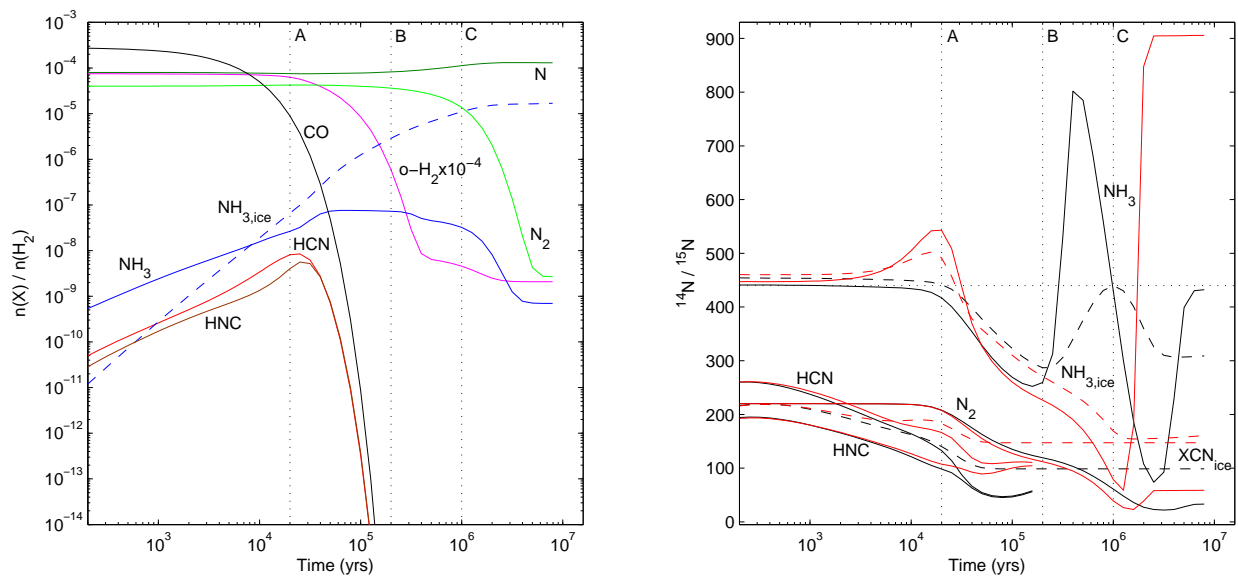} 
\caption{
\textit{Left panel:} Time evolution of the nitrogen chemistry in dense cores, compared to CO and $o$-H$_2$. Crucial times for the $^{15}$N fractionation in nitrogen hydrides are marked by vertical dotted lines, labelled A, B, and C. \textit{Right panel:} Evolution of the $^{14}$N/$^{15}$N ratio in gas-phase molecules and bulk of ammonia and nitrile ices. Black curves represent the current model, while red curves show the corresponding isotopic ratios in the model of \citet{RodgersCharnley08}.
The assumed elemental value of 440 is marked by a dotted horizontal line.
Values for gas-phase nitriles are omitted after $\sim$2$\times$10$^5$~yrs when abundances for both isotopologues are essentially zero due to freeze-out. Dashed curves represent ice-phase species.
}
\label{StandardModFig}
\end{figure*}

Fig.~\ref{StandardModFig} shows the evolution of the major chemical species. The ammonia abundance builds up to about 10$^{-7}$ with respect to H$_2$ around a core age of 10$^5$ years (left panel), which is comparable to what is typically observed \citep{Hotzel04}. Similarly, the peak HCN and HNC agree well with values observed in dense cores \citep{Padovani11}.
At this time the CO is depleted by a factor of $>10^3$. 
A steady-state H$_2$ OPR of 2.3$\times$10$^{-5}$, consistent with attempts to quantify the OPR in the cold dark cloud B68 \citep{Troscompt09}, is reached after about 10$^6$ years.

\begin{figure} 
\includegraphics[width=15cm]{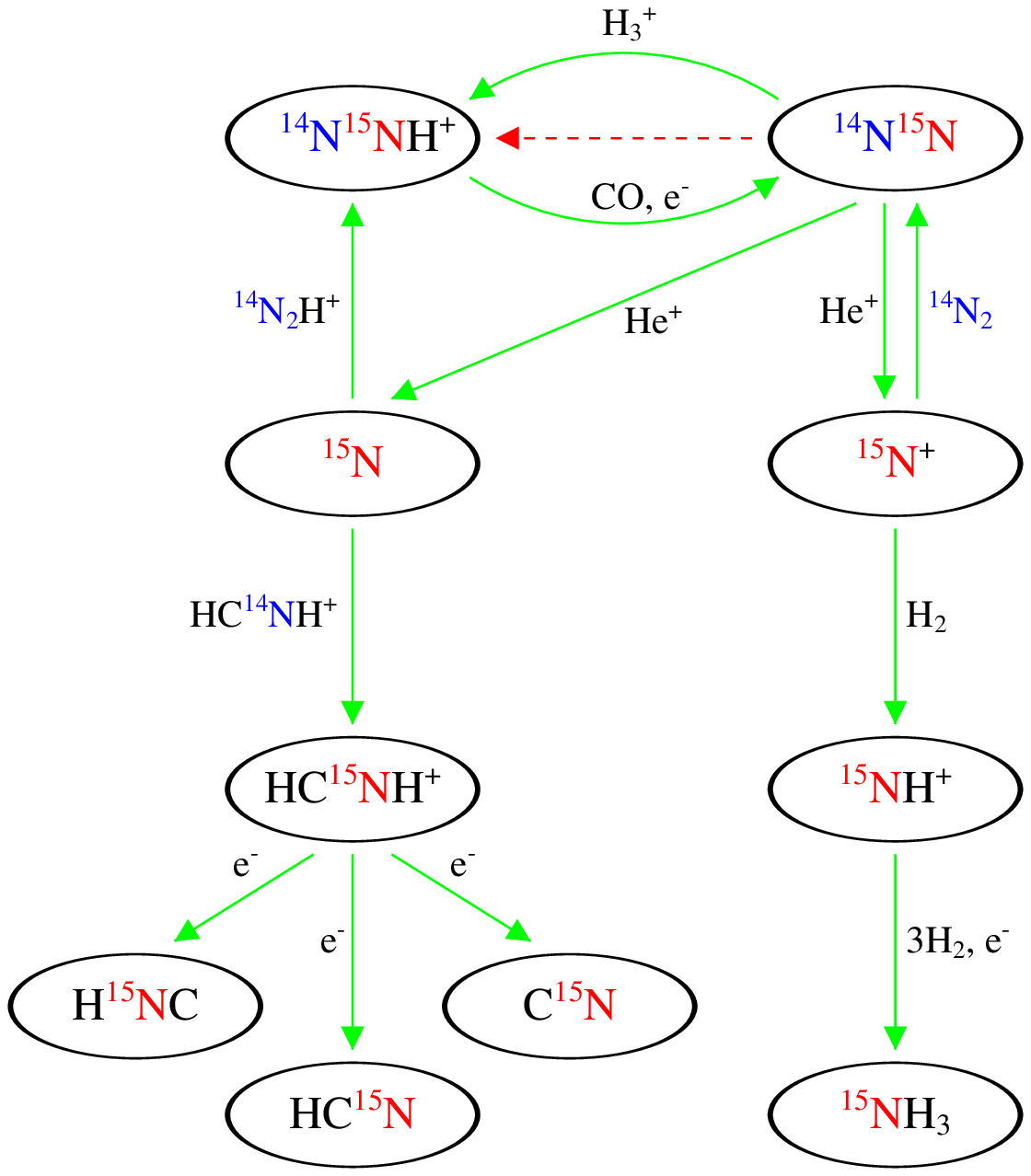} 
\caption{Chemical network showing the main reactions responsible for $^{15}$N enhancement in nitriles and ammonia. }
\label{15NnetworkFig}
\end{figure}

The right panel of Fig.~\ref{StandardModFig} shows the evolution of the molecular $^{14}$N/$^{15}$N ratios in the current model, compared to the corresponding results from the \citet{RodgersCharnley08} model.
As expected (see e.g. Fig.~\ref{15NnetworkFig}), the inclusion of a spin-state dependence does not significantly affect the evolution of fractionation in the nitriles: the $^{15}$N enhancement starts to build up from the initial pool of atomic $^{15}$N early on, and continues to increase until gas-phase abundances have dropped to below 10$^{-13}$. The $^{14}$N/$^{15}$N ratios in gas phase HCN and HNC lie in the range $\approx 50-260$. When all nitriles are frozen out, the $^{15}$N enhancement in the bulk of nitrile ices (CN, HCN, HNC, H$_3$CN) is $\sim$5.


The inclusion of spin-state dependence does however have a dramatic effect on the $^{15}$N fractionation  in ammonia. The relative delay in the timing of the late enhancement peak in the current model, as compared to \citet{RodgersCharnley08}, stems from the higher rate ($\sim 3\times$) of reaction (\ref{HepReac})
(see Sect.~\ref{He+Sect}). 
However, the main difference is that the NH$_3$ enrichment is not monotonic in time. The sharp decrease starting around 2$\times$10$^5$ years (time B) (i.e. an increase in $^{14}$NH$_3$/$^{15}$NH$_3$), is caused by the inefficiency of reaction (\ref{amminitReac}) as the $o$-H$_2$ abundance drops (see Fig.~\ref{StandardModFig} ).


To understand what causes the fractionation patterns, we compare abundances to isotopologue ratios at crucial times in the evolution, marked by vertical lines A, B, and C in Fig.~\ref{StandardModFig}. Initially, the limiting reaction for fractionation is the production of $^{15}$N$^+$, and the $^{14}$N/$^{15}$N ratio in ammonia is simply double that of $^{14}$N$_2$/$^{14}$N$^{15}$N. As gas-phase CO gets depleted after $\sim$10$^4$ years (A), more He$^+$ becomes available to produce $^{15}$N$^+$, which starts to enrich the nitrogen hydrides beyond that of N$_2$.
However, after $\sim$10$^5$ years, the H$_2$ OPR drops below $2\times 10^{-7}$ (B), reaction (\ref{amminitReac}) becomes less efficient, and more  $^{15}$N$^+$ is instead circulated back into molecular nitrogen by
\begin{equation}\label{14N2Reac}
{\rm ^{15}N^+~+~^{14}N_2  ~\rightleftharpoons~\rm{ ^{14}N^+~+~^{14}N^{15}N   }   }  
\end{equation}
\citep{TerzievaHerbst00}, lowering the $^{15}$N fraction in ammonia to less than half the elemental fraction. The reason for ammonia to subsequently become $^{15}$N enriched after 10$^6$ years (C) is that the molecular nitrogen fraction decrease substantially. This both reduces the fraction of $^{15}$N$^+$ converted back into molecular form (Fig.~\ref{15NnetworkFig}), and decreases the amount of $^{14}$N$^+$ available to produce $^{14}$NH$_3$.

Fractionation changes in the bulk of amine ices follows the pattern of gas-phase fractionation, but with a time lag and not as pronounced extreme values. 
While the bulk of amine ices on grains has $^{15}$N fractions within 30\% of the elemental fraction, at any given time the fractionation in the outer monolayer may show much larger variations \citep{RodgersCharnley08}. 

\begin{figure} 
\includegraphics[width=\linewidth]{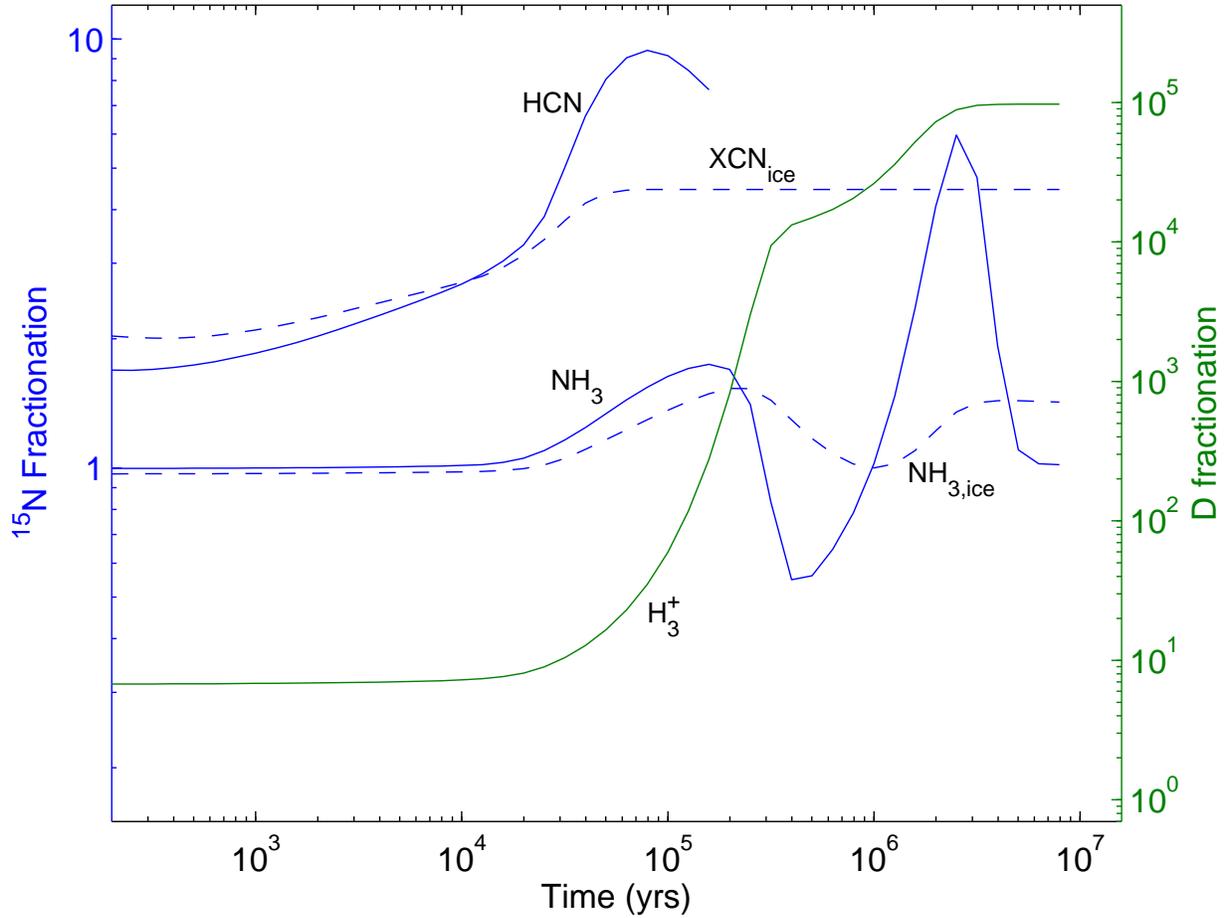} 
\caption{Nitrogen fractionation in ammonia and HCN, both gas-phase (solid) and bulk ice (dashed) in blue curves according to left-hand axis, compared to deuterium fractionation in H$_2$D$^+$/H$_3^+$ (green curve and scale on right hand axis). $^{15}$N enhancements are given relative to the elemental $^{15}$N/$^{14}$N ratio of 1/440, while D enhancements are relative to elemental D/H=1.5$\times 10^{-5}$.}
\label{15NDfig}
\end{figure}

In Fig.~\ref{15NDfig} the calculated  deuterium fractionation in H$_3^+$ is plotted together with the molecular $^{15}$N fractionation. The  D/H ratios  in other gas phase molecules,  in atoms,  and therefore also in the accreted molecular ices, will closely follow that of H$_3^+$. The increase in deuteration follows very closely the drop in $o$-H$_2$ abundance, and after 2$\times$10$^5$ years (time B), ratios of D/H$>$1\% are predicted.  These trends in both deuteration and OPR evolution  agree well with previous investigations  \citep{Pagani11,Flower06}.

Fig.~\ref{15NDfig} shows that gas phase chemistry can indeed reproduce the requisite range of D and $^{15}$N enrichments in nitrile and amine functional groups.  Three fractionation patterns develop as the core evolves: 
 
\begin{trivlist}

\item (I) For  times less than about $\sim  2 \times  10^5$ years, gaseous nitriles can develop large enrichments in $^{15}$N ($\rm HC^{14}N/HC^{15}N \approx 50-260$) but will only be modestly enriched in D, as observed \citep[e.g.  DNC/HNC $\approx 0.01-0.05$;][]{Hirota01};  this will be reflected in the accreted  ices.  
Ammonia will be  similarly enriched in D but less enriched in $^{15}$N  than the nitriles. At this stage the nitrile gas-grain fractionation is complete.

\item (II) Between $\sim  2 \times 10^5 - 10^6$ years,  the remaining ammonia  can become highly fractionated in D but is either not enriched, or even depleted, in $^{15}$N.

\item (III)  Beyond $10^6$ years, ammonia becomes very highly enriched in both D and $^{15}$N.
 
\end{trivlist}


Thus,  the model naturally produces a range of D-$^{15}$N fractionation in nitrile and amine functional groups that is qualitatively  consistent with that found in primitive organic matter. 
 In general,  the highest $^{15}$N enrichments and lowest D enrichments should be carried by  nitriles. 
   Amines could account for the fractionation of material that is highly enriched in D only, as well as the material that is simultaneously  highly enriched in both D and  $^{15}$N. This functional-group dependency has already been noted in analyses of meteoritical samples.  
However, an important caveat is that conversion of nitriles to amines by H and D additions on grain surfaces \citep[e.g. hydrogenation of HCN to CH$_2$NH and CH$_3$NH$_2$,][]{Theule11} has to be negligible.
 {\color{black}  The 30\% enrichment of the bulk ammonia ice cannot  account for all the enhanced ammonia fractionation measured in carbonaceous chondrites \citep[${\rm ^{14}NH_3/^{15}NH_3} \approx 185-256$;][]{PizzarelloWilliams12}. However, we predict  ${\rm ^{14}NH_3/^{15}NH_3} \approx 100$  as the highest gas phase enrichment and this will lead to  higher ${\rm ^{14}NH_3/^{15}NH_3}$ ratios in the uppermost monolayers of the stratified ices  \citep{RodgersCharnley08}.
}


Interstellar gas phase $\rm ^{14}N/^{15}N $ ratios in nitriles have been measured in several Galactic sources  selected for being unlikely to have experienced significant isotopic fractionation \citep{AdandeZiurys12}. For cold dark cores, fewer measurements exist; our model predicts very low   $\rm ^{14}N/^{15}N $ ratios for HCN and HNC. 
For ammonia, the  observed ${\rm ^{14}NH_3/^{15}NH_3}$  ratios in dense protostellar cores \citep[$\approx 340$,][]{Lis10}  are consistent with the modest enrichments which can occur 
within $ 10^5$ years. 
 The measurement of   ${\rm ^{14}NH_3/^{15}NH_3}  > 700 $ in L1544 \citep{Gerin09} is consistent with the  prediction that $^{15}$N nuclei can be underabundant in ammonia. It also  suggests   a  chemical age of $\approx  3 \times  10^5$ years for the  fractionated ammonia gas  (Fig.~\ref{15NDfig} ),  and that it should exist in a  region  of the L1544 core spatially distinct from HCN and HNC \citep[cf.][]{Padovani11}.


\section{Conclusions}

By considering the ortho-para dependence in ion-molecule reactions involving H$_2$,   ammonia and related molecules are expected to exhibit a wide range of fractionation in both $^{15}$N and D, consistent with those found in primitive Solar System organic matter. These include very large enrichments in both D and  $^{15}$N, as found in previous models, but also low  $^{15}$N enrichment or even depletion, coupled with modest to large D/H values.  Nitrile functional groups on the other hand, while being the most likely source of  $^{15}$N hotspots, are not predicted to correlate with extreme D enrichment.  These calculations support the premise that the  $^{15}$N isotopic anomalies found in meteorites were set in functional groups formed in a cold molecular cloud.

The predicted $\rm ^{14}N/^{15}N$ and D/H ratios can account for the limited measurements existing for interstellar cloud cores. To further test and constrain these theoretical models, surveys for key molecules and measurements of relevant  $\rm ^{14}N/^{15}N$  ratios in molecular clouds are required.  
A more extensive analysis of how correlations between deuterium and $^{15}$N enrichment in specific molecules are affected by spin-state-dependent chemistry is beyond the scope of this Letter and will be considered elsewhere  (Wirstr\"om et al. 2012, in preparation). Predictions from these models can be tested by measurements of  various multiply-substituted isotopologues \citep[e.g. DC$^{15}$N, ${\rm ^{15}NH_2D }$,][]{Gerin09}, which will be facilitated with the {\it Atacama Large Millimeter Array} (ALMA).


\acknowledgments
This work was supported by NASA's Origins of Solar Systems Program and the Goddard Center for Astrobiology.



\newcommand{\noopsort}[1]{}

\clearpage
\newpage
	
\newpage



\end{document}